\documentstyle[aps]{revtex}

\newcommand{\rr}[4]{#1, {\it #2 \/}{\bf #3} #4}

\def\al{\alpha}
\def\be{\begin{equation}}
\def\ee{\end{equation}}
\def\bea{\begin{eqnarray}}
\def\eea{\end{eqnarray}}
\def\l{\label}
\def\g{\gamma}
\def\tY{\tilde Y}
\def\nn{\nonumber\\}
\def\d{\delta}

\begin{document}
\title{Conformal invariant saturation}
\author{H. Navelet and R. Peschanski\thanks{%
CEA, Service de Physique Theorique, CE-Saclay, F-91191 Gif-sur-Yvette Cedex,
France}}
\maketitle

\begin{abstract}
We show that, in onium-onium scattering at (very) high energy, a transition to 
saturation happens due to  quantum fluctuations of QCD dipoles. This 
 transition starts when the order $\alpha^2$ correction  of the dipole loop is 
compensated by its faster energy evolution, leading to a negative interference 
with the tree level amplitude. After a derivation of the 
 the one-loop dipole contribution
using conformal invariance of the elastic 4-gluon amplitude  in high energy QCD, 
we obtain an exact expression of the saturation line in the plane (Y,L) where Y 
is  the total rapidity  and L, the logarithm of the onium scale ratio. It shows 
universal features implying  the Balitskyi - Fadin - Kuraev - Lipatov (BFKL) 
evolution kernel and the square of the  QCD triple Pomeron vertex. For large L, 
only the higher BFKL Eigenvalue contributes, leading to a saturation depending  
on leading log perturbative QCD characteristics.  For initial onium scales of 
same order, however, it  involves an unlimited  summation over all conformal 
BFKL Eigenstates. In all cases, conformal invariance is preserved for the 
saturation mechanism based on dipole loops.
\end{abstract}
\bigskip

\section{Saturation from QCD dipoles}
A regime of QCD at small coupling constant where the 
density of partons begins to {\it saturate} has since long been the subject of 
many studies  \cite 
{GLR,Mueller,Nikolaev,Mueller1,Mueller2,Collins,Bartlevin,venugopalan,Salam,Gots
man,Balitsky,Jalilian-Marian,Weigert,Braun,Kovchegov,levtuch,Golec-Biernat,
motyka}. In one class of models, one expects the effect due to the high energy 
multiplication of partons due to the QCD dynamics. Another way which was 
proposed to investigate high density partonic 
effects is 
the consideration of collisions at high energy on heavy nuclei. Indeed, the 
number of partons is supposed to be large in the background of the 
collision providing boundary conditions favorable for the saturation mechanism 
to 
happen.
In the present paper, we will focus on the first scheme, i.e. the saturation 
mechanism due to 
the energy evolution of the parton density in a purely perturbative QCD 
framework. More precisely, we will consider even ``perturbative QCD'' boundary 
conditions, in order to select the features of saturation which could be fully 
calculated from the QCD lagrangian. 

On the experimental ground, saturation has not been clearly identified in the 
evolution of structure functions in the HERA range, even if some evidence for 
partial saturation has been found by analyzing vector meson diffractive 
production in impact parameter space \cite {munier}. The problem of identifying 
saturation effects in actual or future experiments is very stimulating, since 
the increase of the energy range  is at reach at Tevatron and later at 
LHC. Even, some lessons will be available from experiments with heavy nuclei at 
RICH. The goal of our purely perturbative calculations, even if they appear to 
be far from the experimental realizations, is to give an example of saturation 
which could be fully computed, without referring to a non-perturbative ansatz. 
In practice, later on, it will be possible, using factorization properties, to 
combine our calculations with some reasonable non-perturbative wave functions to 
get predictions for realistic processes, like hard hadron-hadron interations at 
Tevatron and LHC.

Our idea is to start from the most elementary hard processes, namely onium-onium 
scattering at (very) high energies.  The reaction can be described 
\cite{Mueller1,Mueller2} by the wave-functions of the onia in terms of $q \bar 
q$ colour singlet dipoles at the time of the (hard) interaction, which then 
interact through double gluon exchanges. This description, which corresponds to 
the tree level dipole interactions, is equivalent to the usual Balitsky, Fadin, 
Kuraev, Lipatov (BFKL) formalism \cite{bfkl} for the same process, as can be 
proven 
\cite{equiv} using the conformal invariant symmetry of the  BFKL evolution 
equation. When energy increases,  the wave functions  has an ever increasing 
density of colour dipoles and the tree level dipole interaction is expected 
to be modified by unitarity corrections which are already 
present  in the perturbative regime. The first order correction is the 
one-loop dipole contribution which was first considered in Ref. \cite 
{nav} as a formal calculation of next-to-next order contributions in the 
coupling constant. In the present paper we show that, despite its higher order,  
this one-loop unitarity correction 
gives rise to the onset of saturation in onium-onium scattering and thus allows 
one to give a complete perturbative QCD description of the saturation critical 
line. This saturation mechanism is to be distinguished from the one which would 
come from the interaction of a onium (e.g. a collection of independent dipoles) 
with a large nucleus (see, e.g. 
\cite{venugopalan,Balitsky,Jalilian-Marian,Weigert,Kovchegov,Golec-Biernat,
motyka}), which, in some cases,  can be associated with a non-linear branching 
process 
without dipole
recombination. In the case we consider, the non-linearities due to the high 
density of partons correspond to the recombination of QCD dipoles.

\eject
We write the onium-onium elastic cross-section 
\be
\frac {d\sigma}{dQ^2} = \frac 1{4\pi}\ \vert F_Q \vert ^2\ ,
\l{cross}
\ee
 where $Q$ is the transfer momentum. The amplitude reads
 \be
 F_Q = -i \int d^2\rho \ d^2\rho'\ \int_0^1 dz\  dz'\ 
\Phi(\rho,z)\Phi(\rho',z')\ f_Q 
(\rho,\rho'|Y)\ ,
 \l{amplitude}
 \ee
where the   onium  is defined as a 
heavy 
$\overline q q$ state  with $q$ (resp. $\overline q$) having the fraction $z$ 
(resp. 
$1\!-\!z$) of the longitudinal momentum   in an infinite momentum frame. It is 
defined by its wave-function $\Phi (\rho,z)$ over a basis of dipole states of 
transverse size $\rho \equiv \rho_0-\rho_1,$ where $ \rho_0,\ \rho_1$ are the 
positions of the dipole's constituants in transverse coordinate space. $f_Q 
(\rho,\rho'|Y)$ is the dipole-dipole elastic amplitude for a full rapidity 
interval $Y$ and a momentum transfer $Q,$ the calculation of which at one 
dipole-loop 
order and saturation properties are the subject of the present paper.

In fact the physical origin of the saturation mechanism that we want to 
investigate is the following: it is known \cite{Mueller1} that the wave function 
of an Onium in the infinite momentum frame develop a large number of dipole 
states as a function of the rapidity (which plays the r\^ole of time in a 
cascading mechanism). However, the larger is the dipole number, the smaller is 
their size and thus they do not contribute to saturation. This can be proven 
rigorously by the fact that the dipole-dipole cross-section add together to give 
back the BFKL cross-section \cite {equiv}.  It is interesting to note the use of 
conformal invariance to get this exact bootstrap property between an 
``s-channel'' and ``t-channel'' description of the same amplitude.

However, now if  one computes the dipole-dipole interaction at one-loop level, 
the system is allowed to take into account quantum fluctuations where virtual 
dipole pairs may be created during the time of interaction. Moreover, all sizes 
of dipoles can contribute in the loop, which goes beyond the limitations of the 
tree-level mechanism. This is expected from perturbative unitarity arguments 
\cite{Mueller1} which expect the damping of the energy dependence of the 
amplitude by saturation effects due to multi-loop dipole contributions. This is 
precisely the goal of our work to put this expectation on a quantitative level.

Through unitarity arguments, the QCD saturation phase corresponding to the 
mechanism we consider requires an estimate of all dipole loop contributions, at 
least in the high energy approximation. In the present paper we will concentrate 
on the first step, i.e. the transition to saturation which is expected to be due 
to the interfernece between the tree and 1-loop level. We expect that the 
techniques which we were led to consider in this work, relying notably on the 
conformal invariance properties, will be useful to investigate this intringing 
phase of matter.

\bigskip

Our main results are the following:

{\bf i)} Up to  one-loop QCD dipole level, the dipole-dipole amplitude reads
\bea
f_Q (\rho,\rho'|Y) &\equiv&  f^{(0)}_Q + f^{(1)}_Q  = 
\frac { \alpha ^2}{4}\ \vert\rho\rho'\vert \int dh 
\ \overline 
E^h_Q 
(\rho') E^h_Q (\rho)\ {d(h)}\ { e}^{\omega(h)Y} \nn
 &{\bf -}& \ \left(\frac {2\alpha}{\pi^4}\right)^4\ \frac {\pi\  
e^{2\omega(\frac12) Y}}{\left[\omega''(\frac12) 
Y\right]^3}
\ \vert \rho\rho^{\prime}\vert \int dh \ (-)^n\ \overline E^h_Q
(\rho^{\prime})  E^h_Q(\rho)\  \left\vert \frac  {g_{3{\cal P%
}} (h,\frac12,\frac12)}{\chi  (h)\!-\!2\chi  (\frac12)}\right\vert ^2 \ .
\label{full}
\eea

In equation (\ref {full}), the symbolic notation
$
\int dh \equiv \sum^{\infty}_{n=-\infty} \ \int d\nu
$ corresponds to the integration over the quantum numbers associated to the 
continuous unitary irreducible representations of $SL(2,{\cal C}),$ namely  
\be
h=i\nu + \frac {1-n}2;\ \tilde h=i\nu + \frac {1+n}2\ ;\ n \in {\cal N}\ ,\ \nu 
\in {\cal R}\ .
\l{h}
\ee
The $SL(2,{\cal C})$ Eigenvalue  
of the BFKL kernel is
\begin {equation}
\omega(h) = \frac { \alpha N_c}{\pi} \ \chi  (h) \equiv 
\frac { \alpha N_c}{\pi}\ 
\left\{2\Psi(1)-\Psi\left(h\right)-\Psi\left(1-\tilde h\right)\right\},
\label{6A2}
\end{equation}
where $\Psi \equiv  (\log\Gamma)^{\prime},$   and $E^h_Q (\rho)$  are the 
$SL(2,{\cal C})$ Eigenfunctions \cite 
{lip}. 
$g_{3{\cal P%
}} (h,h_a,h_b)$ is the known QCD triple Pomeron coupling function 
\cite{pe,triplep}, whose definition is given in Appendix {\bf 1}, together with 
$E^h_Q (\rho)$ and the other ingredients (such as  the functions $a(h)\ , d(h)$) 
of the  
$SL(2,{\cal C})$-invariant formalism. 

The symmetry of the tree level amplitude remains valid at 
one-loop level, leading to a  {\it conformal invariant saturation} 
scheme.

{\bf ii)} The evaluation of formula (\ref{full}) when the condition 
$L 
\equiv\log\left\vert\frac 
{\rho}{\rho'}\right\vert\ \gg 1$ is fulfilled can be done in the forward 
direction ($Q=0$) in 
terms of  the leading Eigenfunction and Eigenvalue of the BFKL kernel, namely 
with $n=0.$ By contrast, for similar dipole sizes ($L={\cal O}(1)$) 
infinitely many Eigenfunctions and Eigenvalues contribute to the loop 
expression, leading to a quite different structure. 

The QCD critical saturation line is  
derived in the  $L\gg 1$ regime and is given  by: 
\be
\tY \left( \frac 
{\chi  (h^*)-\chi  (h_c)}{\chi  (\frac 12)}\right)-2L\left(h_c-h^*\right)\ =
\tY_c+\frac 52\log\tY-\log L\ ,
\l{line}
\ee 
where
$\tY \equiv \omega(1/2) Y$ and $\tY_c \sim {\cal 
O}\left\{\log\left(1/\al^2\right)\right\}$ is a large  constant (formula 
(\ref{constant1}) in the text). The transition to saturation occurs when 
the first member of (\ref{line}) exceeds the second one. Note that using the 
variable $\tilde Y$ means that the rapidity (i.e. energy) dependence of the 
amplitude is quantified  by reference to the BFKL intercept $1+ \omega(1/2).$ 
Hence the overall $\tilde Y, L$ dependence is determined by ``universal'' 
numbers, independent of the coupling constant.

In formula (\ref{line}), there appears two critical values of the effective  
anomalous dimensions $h_c$ for the tree level and $h^*$ for the one-loop 
contribution. They are both defined as solution of   implicit equations
\be
\frac {\chi  '(h_c)}{\chi  (\frac 12)} = \frac {2L}{\tY}\ ;\  \ \ \ \chi  
(h^*)=2\chi  
(1/2) \ .
\l{critical}
\ee
 $h_c$ is a moving  saddle point in the 
tree-level integrand of (\ref{full})
while 
$h^*\sim.183 $ is the fixed location of a double pole singularity in the 
one-loop integrand of (\ref{full}).

The plan of the paper is the 
following. The next section II is devoted to the derivation of formula 
(\ref{full}) 
and  Section III to that of formula (\ref{line}). A disussion of the resulting 
saturation line and an 
outlook are proposed in  section IV. The $SL(2,{\cal C})$-invariant 
formalism which we use is presented in Appendix {\bf 1} and   some 
technical parts of the   derivation of the one-loop amplitude are present in 
Appendix {\bf 2} and {\bf 3}.

\bigskip

\section{The one-loop dipole amplitude}

At one-loop level, the amplitude $f_Q^{(1)}$ has a simple and physically  
appealing formulation \cite{Mueller1}  in the transverse coordinate space, see 
Fig.1. We write
\be
f_Q^{(1)}(\rho,\rho'|Y) \equiv \frac 1{2\pi}\ \int d^2r \ { e}^{iQ}\ 
T^{(1)} 
\left( 
\rho _{0}\rho 
_{1};\rho' _{0}\rho' _{1} | Y\right)\ ,
\l{fourier}
\ee
with impact parameter $r.$ The dipole-dipole amplitude in coordinate space reads
\begin{eqnarray}
& &T^{(1)} \left( \rho _{0}\rho _{1};\rho' _{0}\rho' _{1} | 
Y\!=\!y\!+\!y'\right) \equiv 
{\bf -} \int_0^y d \overline y 
\int_0^{y'} d \overline y' \int \frac{d^{2}\rho _{a_0} 
d^{2}\rho _{a_1}d^{2}\rho _{b_0}d^{2}\rho _{b_1}}{\left| \rho
_{a}\ \rho _{b}\right| ^{2}} \frac{d^{2}\rho _{a'_0}d^{2}\rho _{a'_1}d^{2}\rho 
_{b'_0}d^{2}\rho _{b'_1}}{\left| \rho
_{a'}\ \rho _{b'}\right| ^{2}}\nonumber \\ 
&\times&n_{2}\left( \rho _{0}\rho _{1};\rho _{a_0}\rho
_{a_1},\rho _{b_0}\rho _{b_1} | y\!-\!\overline y,\overline y\right) 
\overline n_{2}\left( \rho' _{0}\rho' _{1};\rho' _{a_0}\rho'
_{a_1},\rho' _{b_0}\rho' _{b_1} | y'\!-\!\overline y',\overline y'\right) \  
T(\rho 
_{a_0}\rho
_{a_1},\rho _{a'_0}\rho
_{a'_1})\ \overline T(\rho _{b_0}\rho _{b_1},\rho _{b'_0}\rho _{b'_1})
\label{18A}
\end{eqnarray}
where  $\rho _{0}, \rho _{1}$   are the transverse 
coordinates of one of the initially colliding dipoles (resp. $\rho' _{0}\ ,\rho' 
_{1}$ for the second one), while  $\rho _{a_0}\rho _{a_1}$ and $\rho 
_{b_0}\rho 
_{b_1},$ are the two interacting dipoles emerging from
the dipole $\rho _{0}\rho _{1}$ after evolution in rapidity (resp. $\rho_{\{i\}} 
\to 
\rho_{\{i'\}}$, for the second one). 

 The  overall minus sign comes from  the well-known 
AGK rules\cite{AGK} relating  the amplitudes to the discontinuity here 
calculated in the c.o.m. frame of the dipole-dipole reaction. 
The tree level interaction amplitudes
$T$ 
are given  in Appendix {\bf 1}.
In formula (\ref{18A}),  $n_{2}\left( \rho _{0}\rho _{1};\rho _{a_0}\rho
_{a_1},\rho _{b_0}\rho _{b_1} | y\!-\!\overline y,\overline y\right)$ stands for 
the probability of producing two dipoles 
after a {\it mixed} 
rapidity evolution \cite{Mueller1,nav}, namely 
with 
a rapidity $y-\overline y$ with one-Pomeron type of evolution and  a rapidity 
$\overline 
y$ with two-Pomeron type of evolution, see Fig.2.  Further on, one  has to 
integrate over $\overline 
y, \overline y'$ . The solution for $n_{2}$  is given in Appendix {\bf 2}.
The 
integration over intermediate variables yields a drastic 
simplification due to the appearance of 
quite a few $\delta$-functions. This tedious but straightforward derivation is 
given in detail in Appendix {\bf 3}. 

\bigskip

Starting with the general formula (\ref{1loop1}) obtained in Appendix {\bf 3},  
we first evaluate the 
high energy behaviour of the integral over $h_{a,b}$
the  dominant contribution of which, as well-known from 
the 
ordinary BFKL analysis, is obtained   for $n_{a,b}=0.$ The integrals
are dominated by the usual BFKL saddle-points at $h_{a,b}=\frac 12;\ 
\omega(\frac 
12)=4\log2.$ 
Taking  into 
account in the saddle-point derivation the logarithmic corrections due to 
the zeroes in the  prefactors at $h_{a,b}=\frac 12,$ one gets:
\be
f^{(1)}_Q\ (\rho,\rho^{\prime}|Y) \sim {\bf -} \left(\frac { 
2}{\pi}\right)^8\ \left(\frac {\alpha}{2\pi^2}\right)^4\ \frac {\pi\  
e^{2\omega(\frac12) Y}}{\left[\omega''(\frac12) 
Y\right]^3}
\ \vert \rho\rho^{\prime}\vert \int dh \ (-)^n\ \overline E^h_Q
(\rho^{\prime})  E^h_Q(\rho)\  \left\vert \frac  {g_{3{\cal P%
}} (h,\frac12,\frac12)}{\chi  (h)\!-\!2\chi  (\frac12)}\right\vert ^2 \ .
\label{full1}
\ee
Formula (\ref{full1}) is then reported in the full amplitude (\ref{full}).
\bigskip

A few comments on our result (\ref{full1}) are in 
order:

i) Conformal invariance is explicitly realized, due to the decomposition on the 
combination of  $SL(2,{\cal C})$ Eigenvectors $ \overline E^h_Q
(\rho^{\prime})  E^h_Q(\rho).$ This property being shared by the tree-level 
amplitude, see 
(\ref{4A}), is thus valid for the whole amplitude (\ref{full}) including 
one-loop 
level. Hence, saturation which is expected from the compensation between tree 
level 
and one-loop due to the different energy dependence is genuinely a conformal 
invariant phenomenon. We conjecture that this will be true to all orders, 
since 
only tensorial 
$SL(2,{\cal C})$ properties play a r\^ole in the symmetry properties of the 
amplitude. Thus the 
saturation regime based on dipole loop contribution, if mathematically 
convergent, is expected to  respect the conformal symmetry.

ii) The key difference between the tree-level and one-loop amplitude properties 
lies in the difference in the analytic singularities of the integrands. While 
the 
tree level term in (\ref{full}) is dominated by a moving saddle-point at high 
energy, 
the 
one loop amplitude (\ref{full1}) has double poles at $\chi  (h)\!= 
2\chi(1/2)=\!8\log2$ 
which 
depend only on the anomalous dimensions $h$ and thus are {\it independent} of 
the 
energy.
This very peculiar property means that the one-loop contribution may depend 
(and 
thus can give information) on BFKL properties which are not reached in the 
conventional analysis. In particular, all higher conformal spins may 
contribute, 
contrary to the tree level contribution. Also, 
contributions 
for values of $h$ not limited to the interval $[0^+,\ \frac 12]$ can give rise 
to 
poles. This may be interpreted as unlimited contributions of higher twist terms 
of 
the BFKL solution. 

iii) Apart the factor $\al ^4,$ which is the natural strength scale of the 
one-loop 
amplitude, there is a dynamical factor connected to the square of the function 
${g_{3{\cal P%
}} (h,\frac12,\frac12)}.$ This function is 
the 
QCD analogue of the triple Pomeron coupling. However, it is not the value 
${g_{3{\cal P%
}} (\frac12,\frac12,\frac12)}$ which is here relevant, as was the case in the 
QCD 
dipole analysis of the triple pomeron coupling in \cite{triplep}, since $h=\frac 
12$ 
is 
not solution of the double pole equation. One could also note that the square 
of 
$g_{3{\cal P}} $ appears, contrary to saturation equations on an extended 
target, where the dependence is linear \cite{Balitsky}.

\bigskip

\section{Saturation critical line}

As discussed in section I, saturation is expected to take place when the 
one-loop amplitude, with coupling of order $\al ^4,$ becomes comparable in 
strength 
with the tree level amplitude of order $\al ^2.$
Formula (\ref{full}) is completely explicit and thus can serve for analyzing 
saturation depending on transverse momentum or, equivalently using the Fourier 
transforms (\ref{8A}), on impact parameter. For sake of simplicity we will 
concentrate on the forward contribution at $Q=0,$ that is (through unitarity) 
for 
the 
total cross-section integrated over impact parameter. In this limit, the 
$SL(2,{\cal C})$ Eigenvectors simplify \cite{lip} to give 
\be
   \overline E^h_Q
(\rho^{\prime}) \ E^h_Q(\rho) \to \left(\frac {\rho}{\rho'}\right)^{h-\frac12}\  
\left( \frac {\overline\rho}{\overline\rho'}\right)^{\tilde h-\frac12}\ .
\l{forward}
\ee
Consequently, we may infer that the key ``order parameter'' for the discussion 
of 
saturation will be the scale invariant ratio of incident dipole sizes $ 
{\rho}/{\rho'},$ in conformity with the 
conformal 
invariance of the amplitude.  This will in turn be determined by the ratio of 
the  
characteristic scales of   the incoming systems which are encoded in 
their corresponding QCD dipole wave functions, see (\ref{amplitude}).

The general solution of  (\ref{full}) is easily 
obtained by a summation over the double pole contributions verifying 
$\chi  (h_i)\!= 
\!2\chi  (\frac12),$ whose location are caracterized by an infinite number of 
values 
$h_i)\!= \!(n_i,\g_i).$ One finds a formal double series
\bea
f^{(1)}_{Q=0}\ (\rho,\rho^{\prime}|Y) &\sim& {-} \left(\frac { 2}{\pi}\right)^6\ 
\left( \frac {\alpha}{\pi^2}\right)^4 \frac {{e^{2\omega(\frac12) 
Y}}}{\left[\omega''(\frac12) 
Y\right]^3}\nonumber\\
&\times&\ \sum_{n_i,\g_i}  (-1)^n\ \left\vert \frac {g_{3{\cal P%
}} (h,\frac 12,\frac 12)}{\chi  '(h_i)}\right\vert^2 
\ \left(\frac {\rho}{\rho'}\right)^{h_i}\  
\left( \frac {\overline\rho}{\overline\rho'}\right)^{\tilde h_i}
\ \left\{L+\frac {\partial \log {\left\vert g_{3{\cal P%
}} (h,\frac 12,\frac 12)\right\vert}}{\partial h}\right\}_{h=h_i}\ ,
\l{poles}
\eea
where, for sake of simplicity,  we used the notation $L 
\equiv\log\left\vert
{\rho}/{\rho'}\right\vert\ .$

For incident   dipoles of comparable  size, $L \sim 
{\cal 
O}(1),$ we may  note that 
the one loop result then depends on an infinite sum over  $(n_i,\g_i)$ which is 
due 
to 
the analytic properties of the function $\chi  (h).$ It even raises the question 
of 
the convergence of the double series (\ref{poles}). On a physical ground, since 
the 
anomalous dimensions $h_i$ have no obvious limitation, this situation can be 
interpreted as due to a 
dependence 
of the one loop amplitude on {\it all} higher twist contributions, including 
those with non 
zero conformal spins $n_i.$ This is a quite strange example where higher twist 
calculations are required to all orders in a purely perturbative framework, in 
striking analogy with actual non perturbative situations.

Leaving this regime for further work, we will here focus on the case where 
only the first pole 
with $h^*=(\g_i^*, n_i^*=0)$ contributes, i.e. where the equivalent of a 
leading twist 
contribution 
is sufficient. This approximation is realized in a region where $L \gg 1$ (or $L 
\ll 1$ by symmetry).

The dominant double pole in (\ref{poles}) such that 
$\chi  (h^*)=2\chi  (\frac 12)=8\log2$ 
is $h^*\equiv (n^*=0,\g^*\sim.183).$ Consequently, factorizing the tree level 
coupling, the resulting one loop 
amplitude can be written
\be
f^{(1)}_{Q=0} (L,\tilde Y) \sim \frac {\al^2}4\ \vert \rho\rho^{\prime}\vert  \ 
\exp\left\{2\tY +2(h^*-\frac 12)L-\tY_1-3\log \tY+\log L \right\}\ ,
\l{loop*}
\ee
where we define $\tY \equiv \omega(\frac 12)Y,$
  and  
\be
\tY_1 = \log \left\{ \frac {\pi^{12}}{ 16 \alpha^2 }\   
\left\vert \frac {\chi  '(h^*)}{ g_{3{\cal P%
}} (h^*,\frac 12,\frac 12)}\right\vert^2 
\ \left(\frac {\chi  (\frac 12)}{\chi  ''(\frac 12)}\right)^3\right\}\ ,
\l{constant}
\ee
 which is   {\it a priori} large, being  in the perturbative domain $\al \ll 1$ 
and 
constant up to $1/L$ correction terms.We define the  saturation critical line  
as the line in 
the $(\tY,L)$ where the tree-level contribution equals the one loop one. It is a 
simple extension to 
the 
conformal invariant configuration of the usual one \cite{GLR}. Beyond this 
line, 
saturation is expected to take place. 

We now make   use of the  
well-known saddle-point  approximation of the tree level amplitude
\be
f^{(0)}_{Q=0} (L,\tilde Y) \sim \frac {\al^2}4\ \vert \rho\rho^{\prime}\vert  \ 
\exp\left\{\frac {\chi(h_c)}{\chi(1/2)} \tY +2(h_c-1/2)L-\log \left(\frac 
1{h_c^2(1\!-\!h_c)^2}\sqrt{\frac 
{2\pi\chi  (\frac 12)}{\chi  ''(\frac 12)}}\right) 
-1/2\log \tY\right\}\ ,
\l{tree*}
\ee
where 
the tree level saddle point $h_c$ is given by the well known implicit 
equation given in  (\ref{critical}).
Inserting  the results  (\ref{loop*}) and (\ref{tree*}) in (\ref{full}) with a 
redefinition of the constant
\be
\tY_c=\tY_1+\log \left(\frac 1{h_c^2(1\!-\!h_c)^2}\sqrt{\frac 
{2\pi\chi  (\frac 12)}{\chi  ''(\frac 12)}} 
\right)\ ,
\l{constant1}
\ee
we get the  equation (\ref{line}) announced in section I for  
the saturation 
critical line.

\section {\bf Discussion, outlook}

Let us describe general features of the critical line, when
$\al$ is very small but otherwise not specified.
Following expression (\ref{line}), it is straightforward to realize that there 
is an absolute bound when $L,Y \gg 1$ given by $\tY/2L > 
\chi  (\frac 12)/\chi  '(h^*)\sim .09.$ Indeed below this bound, the 
compensation 
can no 
more take place between tree level and one loop contributions. It is interesting 
to note that the bound implies that a renormalization group evolution, with $L$ 
moving to large values while keeping $Y$ fixed, cannnot lead from a weak 
density  to a saturation region, as expected from the  general theoretical 
picture of the QCD dynamics of partons. 

One can delimit three typical regimes in the saturation 
transition, depending upon the behaviour of the tree  level amplitude: 
\bigskip

{\bf I} $0 \ll L_0 -> L_1  \ll \tY:$
in this region, saturation is obtained in the region $\tY > 
\tY_0+2L\left(\frac 12-h^*\right) \sim \tY_0+.63L,$ where $\tY_0$ is a large 
effective rapidity scale incorporating the logarithmic prefactors.
\bigskip
 
{\bf II} $L_0 \ll L_1->L_2 < \tY:$ 
It is well-known that the BFKL behaviour discussed in {\bf I} acquires an 
extra ``diffusion term'' (of the gluon transverse momenta along the BFKL ladder 
contributions) in the intermediate region. Within the saddle-point 
approximation it leads to the following saturation curve:
\be
\tY > \tY_0+2L\left(\frac 12-h^*\right) -\frac {\chi  \left(\frac12\right)}{\chi   
''\left(\frac12\right)}\ \frac {2L^2}{\tY}\ ,
\l{intermediate}
\ee 
where the last term is due to diffusion. Quite interestingly, saturation is 
favoured by diffusion. Indeed, formula (\ref{intermediate}) approximately 
interpolates between the behaviour {\bf I} and {\bf II} by a smooth decrease 
of the limiting slope of $\tY$ {\it versus} $L.$

{\bf III} $ L_2 \sim \tY < L:$ near the bound $2L/\tY < 
\chi  (\frac 12)/\chi  '(h^*) \sim 10.$
In this region, one can use  the relations $$h_c \sim
\sqrt {\chi  (\frac 12)\frac {2L}{\tY}}\ ;\ \ \ \  \chi  (h_c) \sim \frac 
1{h_c}\ 
.$$
Using this and again introducing the effective large rapidity scale $\tY_0$ one 
obtains, within a good approximation at large $(\tY,L),$  the constraint 
\be
\tY > \left\{\sqrt {\frac {8L}{\chi  \left(\frac 12\right)}}+{\sqrt {\frac 
{\tY_0}2}}\right\}^2\ .
\l{boundary}
\ee
It is easy to check that (\ref{boundary}) satisfies the bound $\tY/2L > 
\chi  \left(\frac 12\right)/\chi  '(h^*).$
\bigskip

We have not yet discussed the r\^ole of the logarithmic prefactors. Using 
our explicit expressions, it 
will be easy to incorporate them in a more complete study which we leave for 
the future. They may play a phenomenological r\^ole. Indeed, it is important to 
note that the large factor  $\frac 52\log\tY$ of (\ref{line}) is due to the 
double zeroes appearing in 
the loop couplings to the intermediate dipoles, see (\ref{1loop1}). This has 
the important effect of a strong energy dependent damping on the effective 
triple Pomeron couplings, which could otherwise be large \cite{triplep}. 
Another possible important r\^ole would be, by mere extrapolation to higher 
dipole loops, to establish a useful hierarchy between these different orders. 
Indeed the coupling strength expected at order $n,$ namely $\sim (\al^{2} 
\exp{\omega(\frac 12)Y})^n$ cannot provide a hierarchical structure in the 
saturation regime. We are thus conjecturing a series of critical saturation 
lines corresponding to the onset of compensation between a growing number of 
higher loop order contributions. The study of this regime, necessary to 
understand the full saturation behaviour in the conformal invariant set-up will 
be pursued in the near future.

In fact, our work suggests a strategy to attack the full problem of saturation 
from dipole loops. It seems possible to analyze the $n$-loop dipole diagrams in 
the same way as we have shown here for the 2-loop case. It is clear that many 
diagrams can contribute, depending on the multiple rapidities where new dipoles 
can branch. However, we may conjecture that the integration over these 
intermediate rapidity intervals  would  dominantly  lead to the QCD 
multi-Pomeron exchange and the corresponding $1 \to n$ dipole vertices which 
were analyzed in Ref.\cite{janik}. Then, it will be possible to explore the 
conformal invariant saturation regime it self and not only the transition to it.

Another point merits some care, when the onium typical sizes are of the same 
order. A quite interesting regime seems to set in when the sum (\ref{poles}) 
over all the double poles in the one-loop integrals becomes relevant. While 
being still at small coupling constant, this regime shares with the 
non-perturbative one the features of a summation over higher twist 
contributions. It would be nice to solve this case in detail.

Finally, the step between our calculations and actual experimental 
possibilities, which is not done in the present work, could be done either 
through factorization properties or by the consideration of ``hard'' 
subprocesses with very high energies like Mueller-Navelet jets at LHC or 
$\g^*\!-\!\g^*$ collisions at a future linear accelerator. Relevant studies are 
certainly deserved in the future.  

{\bf Acknowledgements}

\noindent We acknowledge fruitful discussions with Yuri Kovchegov, Heribert 
Weigert and Edmond Iancu.

\eject
\setcounter{equation}{0}
\renewcommand{\theequation}{A.\arabic{equation}}

\section*{Appendix 1}
\subsection*{$SL(2,{\cal C})$-invariant formalism}

Using the $SL(2,{\cal C})$-invariant formalism, the solution of the BFKL 
equation obtained at leading log level of perturbative QCD \cite {bfkl} reads
\begin {equation}
f^{(0)}_Q (\rho,\rho'|Y)=\frac { \alpha ^2}{4}\ \vert\rho\rho'\vert \int dh 
\ \overline 
E^h_Q 
(\rho') E^h_Q (\rho)\ {d(h)}\ {e}^{\omega(h)Y},
\label{4A}
\end{equation}
where the $SU(3)$ coupling$ \alpha$  is fixed, small but unspecified, in the 
leading log approximation we consider. 

In equation (\ref {4A}), the symbolic notation
$
\int dh \equiv \sum^{\infty}_{n=-\infty} \ \int d\nu
$ corresponds to the integration over the quantum numbers associated to the 
continuous unitary irreducible representations of $SL(2,{\cal C}),$ namely  
\be
h=i\nu + \frac {1-n}2;\ \tilde h=i\nu + \frac {1+n}2\ ;\ n \in {\cal N}\ ,\ \nu 
\in {\cal R}\ .
\l{h1}
\ee
$E^h_Q (\rho)$ and $\omega(h)$ are, respectively, the 
$SL(2,{\cal C})$ Eigenfunctions and Eigenvalues of the BFKL kernel \cite 
{lip}. 
The Eigenvalues read
\begin {equation}
\omega(h) = \frac { \alpha N_c}{\pi} \ \chi  (h) \equiv 
\frac { \alpha N_c}{\pi}\ 
\left\{2\Psi(1)-\Psi\left(h\right)-\Psi\left(1-\tilde h\right)\right\},
\label{6A1}
\end{equation}
where $\Psi \equiv  (\log\Gamma)^{\prime}.$ The $SL(2,{\cal C})$ Eigenvectors 
are defined\footnote{Note that 
an analytic expression of  the Eigenvectors $E^h_Q (\rho)$ in the mixed 
representation has been provided \cite {equiv} in terms of a combination of 
products of two Bessel functions. For simplicity, we did not include the 
impact 
factors \cite {lip,vertex}. Phenomenologically, the leading contribution to 
the 
amplitude (\ref {4A}) comes from the $n=0$ component which corresponds to the 
BFKL 
Pomeron.} by
\begin {equation}
 E^h_Q (\rho) = \frac {2\pi^2}{\vert \rho \vert \ b(h)}\int d^2r\ {e}^{iQ 
b}\ 
 E^h \left(r\!-\!\frac{\rho}2,r\!+\!\frac{\rho}2\right), 
\label{8A}
\end{equation}
with
\begin {equation}
 E^h \left(r-\frac{\rho}2,r+\frac {\rho}2 \right)\equiv \frac {\vert \rho\vert 
\ b(h)}{8\pi^4}\ \int d^2r\ {e}^{-iQ r}\ E^h_Q (\rho)
 =(-)^{h-\tilde h}
\left(\frac{\rho}{r^2-\frac{\rho ^2}4}\right)^h\  \left(\frac{\overline 
\rho}{\overline 
r^2-\frac{\overline \rho ^2}4}\right)^{\tilde h}, 
\label{10A}
\end{equation}
where $\tilde h = 1-\overline h,$  $r$ is the 2-d impact-parameter, and 
introducing 
useful notations 
\begin {equation}
d(h) = \frac 1{\left(\nu ^2\! +\! \frac {(n\!-\!1)^2}4\right) \left(\nu ^2 
\!+\! 
\frac 
{(n\!+\!1)^2}4\right)}\ ;\  b(h)=\frac {\pi^3 4^{h\!+\!\tilde h 
-1}}{\frac 12 
-h}\ \frac {\Gamma(-i\nu \!+\! \frac {1\!+\!\vert n\vert}2)}{\Gamma(i\nu \!+\! 
\frac {1\!+\!\vert n\vert}2)}\ \frac{\Gamma(i\nu \!+\! \frac {\vert 
n\vert}2)}{\Gamma(-i\nu \!+\! \frac {\vert n\vert}2)};\ a(h)=\frac {\vert 
b(h)\vert^2}{2\pi^2}=\frac {\pi^4/2}{\nu^2\! +\!\frac {n^2}4}\ .
\label{12A}
\end{equation}

We will also make use of the formalism in the full coordinate space,  the 
4-diple amplitude $T^{(0)}$ being defined at tree level  of QCD 
dipole contributions by
\be
T^{(0)}(\rho_0, \rho_1; \rho'_0, \rho'_1|Y) \equiv \frac 1 {2\pi} \int d^2Q \ 
{e}^{-irQ} 
\ f^{(0)}_Q (\rho,\rho'|Y)\ ,
\l{coordinate}
\ee
where the impact parameter is $r=\frac 12 (\rho'_0\! +\! \rho'_1\!-\! \rho_0\!- 
\!\rho_1).$ 

Inserting  (\ref{4A}) into (\ref{coordinate}), one finds
\be
T^{(0)}(\rho_0, \rho_1; \rho'_0, \rho'_1|Y) = 
\frac {\alpha ^2}{4}\  \int dh \ \ G^h(\rho_0, \rho_1; \rho'_0, \rho'_1) 
\left(\nu^2\!+\!\frac {n^2}4\right)\ {d(h)}\ {e}^{\omega(h)Y}\ ,
\l{amplicoord}
\ee
where , using the  $SL(2,{\cal C})$ Eigenvectors in coordinate space 
(\ref{10A}), the Green function  $G^h$ in coordinate space \cite{equiv,lip} can 
be written as
\be
G^h(\rho_0, \rho_1; \rho'_0, \rho'_1) \equiv \int d^2\rho_{\al} \overline 
E^h(\rho_{0'\al},\rho_{1'\al})\ E^h(\rho_{0\al},\rho_{1\al}) = \frac {\vert 
b(h)\vert^2}{\pi^4}\ \vert \rho \rho'\vert\int d^2Q {e}^{-ibQ}\ \overline 
E^h_Q 
(\rho') E^h_Q (\rho)\ .
\l{green}
\ee  
We will also define for convenience  the ``instantaneous'' (i.e. Y=0) amplitude
\be
T(\rho_0, \rho_1; 
\rho'_0, \rho'_1) \equiv T^{(0)}(\rho_0, \rho_1; \rho'_0, \rho'_1|Y=0)\ .
\l{simple}
\ee
Due to completeness and frame independence of the formalism, many formulae can 
be simplified using these ``instantaneous''
amplitudes between dipoles.

The single multiplicity distribution of dipoles $n_{1}$ after an evolution 
characterized by the rapidity $Y$ is simply related \cite{equiv} to the 
previous expressions (e.g. (\ref{green})), namely
\be
n_1(\rho_0, \rho_1; \rho'_0, \rho'_1|Y) = \frac 1{\pi^2\vert \rho'\vert^2}
\int dh \ \ G^h(\rho_0, \rho_1; \rho'_0, \rho'_1) \left(\nu^2+ 
{n^2}/4\right) {e}^{\omega(h)Y}\ .
\l{multicoord}
\ee

Let us finally note a useful completeness formula \cite {equiv} valid at tree 
level:
\bea
T^{(0)}\left( \rho _{0}\rho _{1};\rho' _{0}\rho' _{1} | 
Y\!=\!y\!+\!y_a\!+\!y'\right) &\equiv& \frac 1{2^2(2\pi)^4}\ 
\int \frac{d^{2}\rho _{a_0} 
d^{2}\rho _{a_1}}{\left| \rho
_{a}\right| ^{2}} \frac{d^{2}\rho _{a'_0}d^{2}\rho _{a'_1}}{\left| 
\rho
_{a'}\right| ^{2}} \nonumber\\
&\times& \ n_{1}\left( \rho _{0}\rho _{1};\rho _{a_0}\rho
_{a_1} | y\right)\  
\overline n_{1}\left( \rho' _{0}\rho' _{1};\rho' _{a_0}\rho'
_{a_1}| y'\right)\  T^{(0)}(\rho _{a_0}\rho
_{a_1},\rho _{a'_0}\rho
_{a'_1}|y_a)\ ,
\l{equiv}
\eea
where $\rho _{0},\rho _{1}$   are the transverse 
coordinates of  one initial  dipole (resp. $\rho' _{0}\rho' 
_{1}$ for the second one), $\rho _{a_0},\rho _{a_1}$ are those of the  
interacting dipole emerging from
the dipole $\rho _{0},\rho _{1}$ after evolution in rapidity $y$ (resp. $\rho 
_{a'_0},\rho _{a'_1}$ and $y'$ for the second one),  which interaction is  
described by the amplitude $T(\rho_{a_0}, \rho_{a_1}; \rho_{a'_0}, 
\rho_{a'_1}|y_a)$ as in (\ref{amplicoord}).

Equation (\ref{equiv})  has the physical interpretation that, at tree dipole 
level, the multiplication of dipoles due to rapidity evolution does not lead to 
saturation, but to the same tree level amplitude in a kind of bootstrap 
property.

\eject

\section*{Appendix 2}
\subsection*{The two-dipole density  distribution  $n_{2}$}

The two-dipole density  distribution $n_{2}$ results from the solution of a 
specific  
evolution equation which consists of an extension of the mixed evolution of the 
one formulated in 
ref. \cite{Mueller1,n2,pe}, namely:
\bea
&&n_{2}\left( \left. \rho _{0}\rho _{1};\rho _{a_0}\rho _{a_1},\rho _{b_0}\rho
_{b_1}\right|y\!-\!\overline y,\overline y\right) = \nonumber\\
&&\ \ \ \ =\frac { \alpha N_c}{2\pi^2}\ \int 
\frac 
{dh dh_a dh_b}{\vert \rho_a\rho_b\vert^2}\ \int d\omega_1 \frac {{\bf 
e}^{\omega_1 
(y-\overline y)}}
{\omega_1-\omega \left( h\right)}
\int d\omega \frac {{e}^{\omega y}}
{\omega \left(
h_{a}\right) +\omega \left( h_{b}\right) -\omega }\ 
n_{2}^{h,h_a,h_b} \left(\rho _{0}\rho _{1};\rho _{a_0}\rho _{a_1},\rho 
_{b_0}\rho
_{b_1}\right),
\label{20A}
\eea
 where, by convention, $d\omega_i \to d\omega_i/2\pi$ and
\bea
n_{2}^{h,h_a,h_b} \left(\rho _{0}\rho _{1};\rho _{a_0}\rho _{a_1},\rho 
_{b_0}\rho
_{b_1}\right)= \frac 1{a(h)a(h_{a})a(h_{b})}&\ &\int d^{2}\rho _\alpha 
d^{2}\rho 
_\beta \ d^{2}\rho _\gamma\ \ {\overline {\cal 
R}}^{h,h_a,h_b}_{\alpha,\beta,\gamma} \times \nonumber \\
&\times& \ {E}^{h_{a}}{\left( \rho
_{a_0 \alpha} ,\rho _{a_1\alpha} \right) }\ {E}^{h_{b}}{\left(
\rho _{b_0\beta} ,\rho _{b_1\beta} \right) }\ {E}^{h}{\left( \rho
_{0 \gamma} ,\rho _{1\gamma} \right) }\ ,
\label{21A}
\eea
with
\begin{equation}
{\cal R}^{h,h_a,h_b}_{\alpha,\beta,\gamma}\equiv
\int \frac {d^{2}r_{0}d^{2}r_{1}d^{2}r_{2}}{\left| r_{01}\ r_{02}\ 
r_{12}\right| 
^{2}}
\ E^{h}{\left( r_{0\gamma},r_{1\gamma}\right)}E^{h_{a}}{\left( r_{0\alpha 
},r_{2\alpha }\right)
}E^{h_{b}}{\left( r_{1\beta} ,r_{2\beta} \right)
}
\label{22A}
\end{equation}
where $\rho =\rho _{0}\!-\!\rho _{1},\rho _{a}=\rho _{a_0}\!-\!\rho _{a_1},\ 
\rho _{b}=\rho _{b_0}\!-\!\rho _{b_1}.$

Conformal invariance implies the well-known tensorial reduction \cite{polya}
\be
{\cal R}^{h,h_a,h_b}_{\alpha,\beta,\gamma}\equiv
\left[\rho_{\alpha\beta}\right]^{h\!-\!h_a\!-\!h_b}
\left[\rho_{\beta\gamma}\right]^{h_a\!-\!h_b\!-\!h}
\left[\rho_{\gamma\alpha}\right]^{h_b\!-\!h_a\!-\!h}
\left[\overline\rho_{\alpha\beta}\right]^{\tilde h\!-\!\tilde h_a\!-\!\tilde 
h_b}
\left[\overline \rho_{\beta\gamma}\right]^{\tilde h_a\!-\!\tilde h_b\!-\!\tilde 
h}
\left[\overline\rho_{\gamma\alpha}\right]^{\tilde h_b\!-\!\tilde h_a\!-\!\tilde 
h}\ 
g_{3{\cal P}}\left(h,h_a,h_b\right),
\label{24A}
\ee
where $g_{3{\cal P}}\left(h,h_a,h_b\right)$ happens to be \cite{triplep} the 
QCD triple Pomeron
coupling as obtained in the dipole formulation,
namely:
\be
g_{3{\cal P}}\left(h,h_a,h_b\right)=
\int \frac {d^{2}r_{0}d^{2}r_{1}d^{2}r_{2}}{\left| r_{01}\ r_{02}\ 
r_{12}\right| 
^{2}}
\left[r_{01}\right]^{h}
\left[\frac {r_{02}}{r_0r_2}\right]^{h_a}\ 
\left[\frac 
{r_{12}}{\left(1-r_1\right)\left(1-r_2\right)}\right]^{h_b}\left[\overline 
r_{01}\right]^{\tilde h}
\left[\frac {\overline r_{02}}{\overline r_0\overline r_2}\right]^{\tilde h_a}
\left[\frac {\overline r_{12}}{\left(1-\overline r_1\right)
\left(1-\overline r_2\right)}\right]^{\tilde h_b}.
\label{26A}
\ee

Considering the Fourier transforms (\ref {8A}) for the $SL(2,{\cal C})$ 
Eigenvectors $E^h_q$ in the mixed representation, one writes
\begin{eqnarray}
n_{2}^{h,h_a,h_b} \left(\rho _{0}\rho _{1};\rho _{a_0}\rho _{a_1},\rho 
_{b_0}\rho
_{b_1}\right)= g_{3{\cal P}}\left(h,h_a,h_b\right) \ \frac 
{b(h)b(h_a)b(h_b)}{2^7\pi^{10}}
\vert \rho \rho_a 
\rho_b\vert\ \int d^2q_a d^2q_b d^2Q\ \delta^{(2)}(Q\!+\!q_a\!+\!q_b)
\nonumber\\
 \times \ E^{h}_{Q}(\rho)  E^{h_a}_{q_a} (\rho_a) E^{h_b}_{q_b} (\rho_b) 
\ {e}^{-i\left(
q_{a}r_a + q_{b}r_b + Qr\right)}
\int d^2v\ d^2w \ {e}^{-i\left(
(q_a\!-\!q_b)\frac v2 + Q\frac w2\right)}
\nonumber\\
\times \ 
\left\{\left[v\right]^{-1\!-\!h\!-\!h_a\!-\!h_b}\left[(w\!-\!v)/2\right]^{-
1
+h\!-\!h_a\!+\!h_b}
\left[(w+v)/2\right]^{-1\!+\!h\!+\!h_a\!-\!h_b}\right\}\times
\left\{a.h.\right\},
\label{27A}
\end{eqnarray}
where $\rho_{\alpha\beta}=v,$ $\rho_{\alpha\gamma}+\rho_{\beta\gamma} =w,$ 
$2r_a= \rho_{a0}+ \rho_{a1}, $  $2r_b= \rho_{b0}+ \rho_{1} $ and $r$ is
the overall impact parameter. The notation $\left\{a.h.\right\}$ indicates the 
{\it anti-holomorphic} part of the bracketed term in the integrand for which 
the 
integration variables are complex conjugates and the exponents $h_i$  are 
replaced by $\tilde h_i.$
Equivalently, the distribution $n_2$ for the lower vertex is given by the same 
equation (\ref{27A}) by using prime indices.
\eject

\section*{Appendix 3}
\subsection*{The general  one-loop amplitude}

Implementing equations(\ref{amplicoord}) and (\ref{27A}) in the master equation 
(\ref{full}) and first integrating over impact parameters,
one gets:
$$
(2\pi)^8\ \delta(q_a-q"_a)\delta(q_b-q"_b)\delta(q'_a-q"_a)\delta(q'_b-q"_b)\ .
$$
Then, integrating over the intermediate dipole sizes, and using the 
orthogonality \cite{lip} relation
\be
\frac 1{4\pi^2} \int \frac {d^2\rho}{\vert\rho\vert^2} E^h_q (\rho) \overline 
E^{h'}_q 
(\rho)
= \delta^{(2)}(h,h') + \delta^{(2)}(h,1-h') \   {e}^{i\phi(h)}\ ,
\label{ortho}
\ee
where $  {e}^{i\phi(h)} \equiv E^h_q (\rho)/E^{1\!-\!h}_q 
(\rho)$ is just a $\rho,q$-independent  phase factor given in 
Ref.\cite {lip}, 
one finds
$$
(2\pi)^8\ 2^4\ 
\delta^{(2)}(h_a,h"_a)\delta^{(2)}(h_b,h"_b)\delta^{(2)}(h'_a,1\!-\!h"_a)\delta
^
{(
2)
}(h'_b,1\!-\!h"_b),
$$
where 
 $
\delta^{(2)}(h,h') \equiv \delta_{nn'}\delta(\nu\!-\!\nu').$ The integration 
over 
$q_a\!-\!q_b$ finally gives $(2\pi)^2\ \delta(v\!-\!v').$

Plugging in  formulae (\ref{fourier},\ref{18A}) the result obtained in 
formula (\ref{27A}) and integrating over $\delta$-fuctions, 
one 
gets:
\begin{eqnarray}
f^{(1)}_{Q} \left(\rho,\rho^{\prime}|Y=y\!+\!y'\right) 
= -\ \frac {\alpha ^4\pi^2}8 \left(\frac {\alpha N_c}{%
2\pi^2}\right)^2  \int dh_a dh_b \frac {d(h_a)}{\vert 
b(h_a)\vert^2} \ \frac {d(h_b)}{\vert b(h_b)\vert^2} \nonumber \\
\times \  \int \! dh dh^{\prime}\ \frac {g_{3{\cal P}}\left(h,h_a,h_b\right)}
{ b(1\!-\!h)}\ \frac{
g_{3{\cal P}}%
^*(h^{\prime},h_a,h_b)}{b(h')} \ \vert \rho \rho' \vert \ E^h_Q(\rho)\ 
\overline E^{h'}_Q(\rho')\ {\cal H}_Q(h,h') \nonumber \\
\times \ 
\int_0^y\! d\overline y \int_0^{Y-y'}\! d\overline y^{\prime}\int\! d\omega 
d\omega_1
d\omega^{\prime}d\omega^{\prime}_1 \frac {{e}^{\omega \overline y + 
\omega_1 
(y-%
\overline y)}} 
{(\omega(h_a)\!+\!\omega(h_b)\!-\!\omega)(\omega_1\!-\!\omega(h))}
\ \frac {{e}^{\omega^{\prime}\overline y^{\prime}+
\omega^{\prime}_1 (y^{\prime}-\overline y^{\prime})}}{(\omega(h_a)\!+\!%
\omega(h_b)\!-\!\omega^{\prime})(\omega^{\prime}_1\!-\!\omega(h^{\prime}))}\
, \label{28A}
\end{eqnarray}
where, after the change of variables $w=v(1-2t),\ w'=v(1-2t'),$ 
\be
{\cal H}_Q(h,h')=\int d^2vd^2td^2t'\ {e}^{iQv(t-t')}\ 
\left\{v^{h\!-\!h'\!-\!1}t^{-\!1\!+\!h\!+\!h_b\!-\!h_a} 
(1-t)^{-\!1\!+\!h\!-\!h_b\!+\!h_a}t'^{h_a\!-\!h_b\!-\!h'}
(1-t')^
{-\!h_a\!+\! h_b\!-\!h'}\right\}\ \times\ {a.h.}\ .
\label{30A}
\ee
After integration over $v$ one obtains \cite{geronimo}
\bea
{\cal H}_Q(h,h')&=& \left(\frac 2{\overline Q}\right)^{h-h'}\left(\frac 
2{Q}\right)^{\tilde h-\tilde h'}\ {e}^{i\frac {\pi}2(n-n')}\ \g(h-h')
\int d^2t\ d^2t'\ (t-t')^{h'-h}\nonumber \\ &\times& 
\left\{t^{-1+h+h_b-h_a}(1-t)^{-1+h-h_b+h_a}t'^{h_a-h_b-h'}(1-t')^{-h_a+h_b-h'}
\right\}\times\{a.h.\}\ .
\label{31A}
\eea
where, by definition,  $\gamma (z) \equiv  {\Gamma (z)}/{\Gamma (1-\tilde 
z)}\ .$

The remaining integral can be proven\footnote {One shows using general methods 
\cite{geronimo} that ${\cal H}_Q(h,h')$ is a combination of $\delta$ functions 
whose coefficients can then be computed directly on formula (\ref{31A}).} to 
reduce to 
\be
{\cal H}_Q(h,h')= \left(\frac 2{\overline Q}\right)^{h\!-\!h'}\!\left(\frac 
2{Q}\right)^{\tilde h\!-\!\tilde h'} 
 \frac {\pi^2\ }{(2h\!-\!1)(2\tilde h\!-\!1)} 
\left[\d(h\!-\!h')\!+\!(-)^{n\!+\!n_a\!+\!n_b}
\   \gamma(h\!+\!h_a\!-\!h_b)\ 
\gamma(1\!-\!h'\!-\!h_a\!+\!h_b)\ \d(1\!-\!h\!-\!h')\right]\ .
\label{32A}
\ee
The integral over $h'$ can now easily be performed. The remaining poles at 
$h'=h$ and $h'=1-h$ give twice the same contribution due to the over 
completeness relation of the $E_Q^h$ generators \cite {lip}. One finally 
obtains:

\begin{eqnarray}
f^{(1)}_{Q} (\rho,\rho^{\prime}|Y) =  -\ \frac {\alpha ^4\pi^4}{8}\ \left(\frac 
{\alpha N_c}{%
2\pi^2}\right)^2\int \! dh dh_a dh_b  \left\vert \frac {g_{3{\cal P}}
(h,h_a,h_b)} {b(h_a)b(h_b)b(h)} \right\vert^2  d(h_a) d(h_b) \frac {
(-1)^{n\!+\!n_a\!+\!n_b}}{\vert\frac 12\!-\!h\vert^2 }\ \vert 
\rho \rho' 
\vert \  E^h_Q(\rho)\ 
\overline E^{h}_Q(\rho')\nonumber \\
\times\ \int_0^y \!d\overline y \int_0^{Y\!-\!y}\! d\overline 
y^{\prime}\int\! 
d\omega
d\omega_1 d\omega^{\prime}d\omega^{\prime}_1 \ \frac {{e}^{\omega \overline 
y +
\omega_1 (y-\overline y)}} 
{(\omega(h_a)\!+\!\omega(h_b)\!-\!\omega)(\omega_1\!-%
\!\omega(h))} \ \frac {{e}^{\omega^{\prime}\overline y^{\prime}+
\omega^{\prime}_1 (y^{\prime}-\overline y^{\prime})}}{(\omega(h_a)\!+\!%
\omega(h_b)\!-\!\omega^{\prime})(\omega^{\prime}_1\!-\!\omega(h))}\ .
\label{36A}
\end{eqnarray}

This one-loop amplitude preserves the global conformal invariance of the 
tree-level BFKL 4-gluon amplitude, since the only scale-dependence on $Q$ is 
present in the conformal Eigenvectors
$\vert \rho \rho' \vert \ E^h_Q(\rho)\ 
\overline E^{h}_Q(\rho').$

It was noted in Ref.\cite{nav} that
the integration over rapidity variables yields two different contributions
depending on the sign of the quantity $\omega(h_a)\!+\!\omega(h_b)\!-\!%
\omega(h).$ Indeed for $\omega(h_a)\!+\!\omega(h_b)\!<\!\omega(h),$ the
relevant poles are situated at $\omega\!=\!\omega_1\!=\!\omega^{\prime}\!=\!%
\omega^{\prime}_1 =\omega(h),$ leading to  a contribution
associated with the single Pomeron dependence ${e}^{\omega(h)\ Y}.$ In the
opposite case, namely $\omega(h_a)\!+\!\omega(h_b)\!>\!\omega(h),$ the
relevant poles are situated at $\omega\!=\!\omega_1\!=\!\omega^{\prime}\!=\!%
\omega^{\prime}_1 =\omega(h_a)\!+\!\omega(h_b).$ The resulting amplitude
\begin{eqnarray}
f^{(1)}_Q\ (\rho,\rho^{\prime}|Y) &\sim& {\bf -}\ \frac 1{2\pi^8} \left(\frac { 
\alpha}{2\pi^2}\right)^4 
\vert \rho\rho^{\prime}\vert\ \int dh\   (-1)^n  \ \overline E^h_Q
(\rho^{\prime})  E^h_Q(\rho)\nonumber \\
& & \times\  \int \int dh_a dh_b\ \left[\frac 
{(\frac 12\!-\!h_a)(\frac 
12\!-\!h_b)}{h_a(1\!-\!h_a)h_b(1\!-\!h_b)}\right]^2\ {
e^{\left(\omega(h_a)+\omega(h_b)\right) Y}}\ \left\vert \frac  {g_{3{\cal P%
}} (h,h_a,h_b)}{\chi  
(h)\!-\!\chi  (h_a)\!-\!\chi  (h_b)}\right\vert^2 \ ,
\label{1loop1}
\eea
is the 
one we are interested in for the saturation problem since it possesses
 the QCD ``double Pomeron'' energy behaviour 
${e}^{(\omega(h_a)\!+\!\omega(h_b)) Y} > {e}^{\omega(h)\ Y}.$ This is 
precisely the energy behaviour which will  compensate for 
the higher 
order  $\al^2$ in the coupling constant. Notice that the
expression depends only on the sum $Y=y+y^{\prime},$ as it should do from
longitudinal boost invariance.

\bigskip
 \eject

 \eject

{\bf FIGURES}

\bigskip \input epsf \vsize=8.truecm \hsize=10.truecm \epsfxsize=8.cm{%
\centerline{\epsfbox{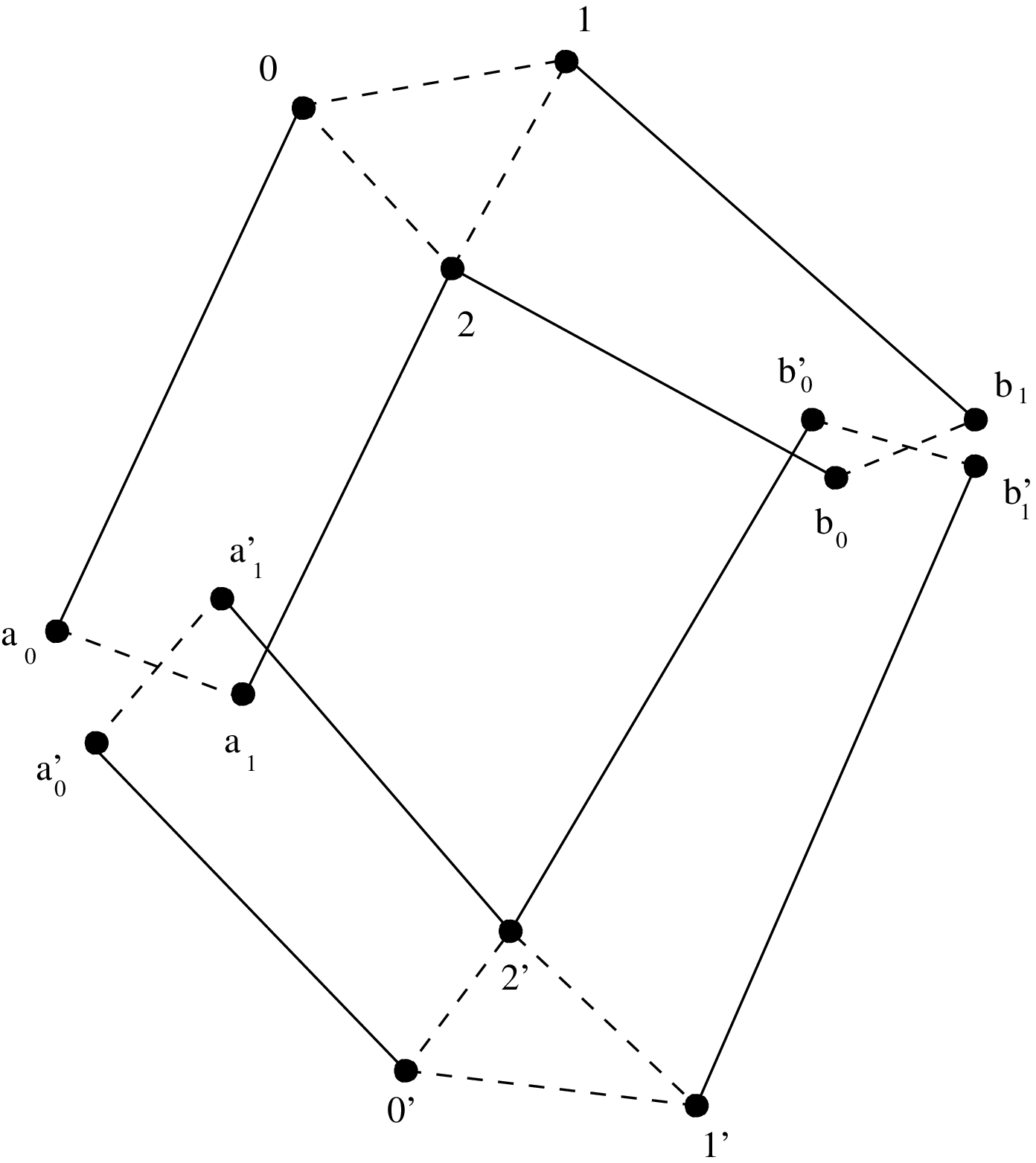}}} {\bf Figure 1}

{\it One-loop dipole amplitude in transverse coordinate space}

The transverse coordinates $\rho _{i}$ of the incident and of the two 
interacting dipoles  are simply denoted by their respective indices ${i}.$ The 
coordinates ${2},{2'}$ refer to the integration points, see Appendix {\bf 
2}.
\bigskip  

\input epsf \vsize=8.truecm \hsize=10.truecm \epsfxsize=8.cm{%
\centerline{\epsfbox{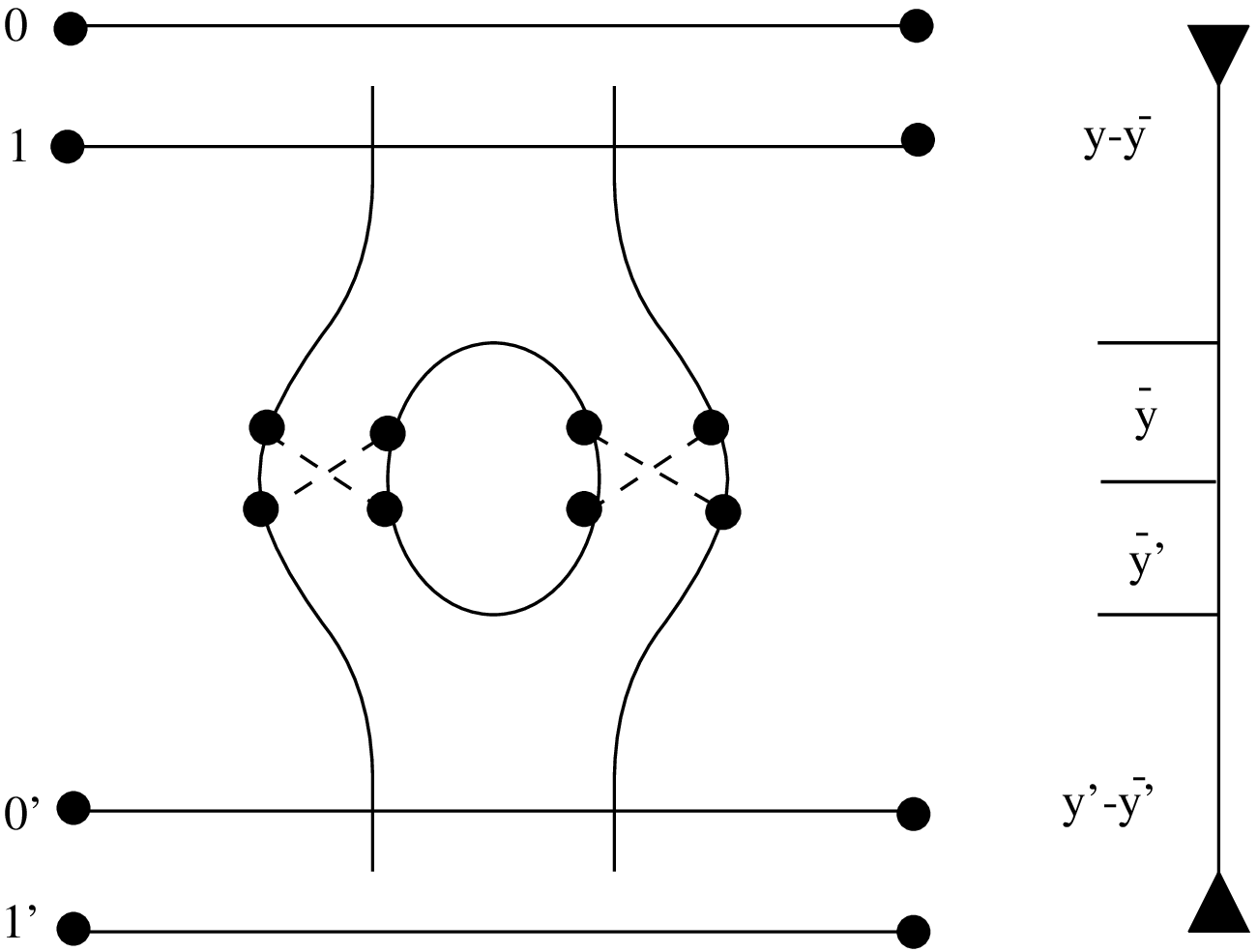}}}{\bf Figure 2} 

{\it One-loop dipole amplitude in rapidity space}

The initial
dipoles $\rho _{0}\rho _{1}, \rho^{\prime}_{0}\rho^{\prime}_{1}$ evolve in the 
rapidity range $y-\overline y^{\prime},y^{\prime}-\overline y^{\prime}$ with 
one-Pomeron type
of evolution and then with two-Pomeron type of evolution over the range  
$\overline y,
 \overline y^{\prime}.$ Note that $y+y^{\prime}\equiv Y,$ is the total 
rapidity range. After integration over $\overline y, \overline y^{\prime},$ see 
Appendix {\bf 3}, 
the
resulting amplitude depends on the total range $Y$. 

\bigskip 

\input epsf \vsize=12.truecm \hsize=14.truecm \epsfxsize=12.cm{%
\centerline{\epsfbox{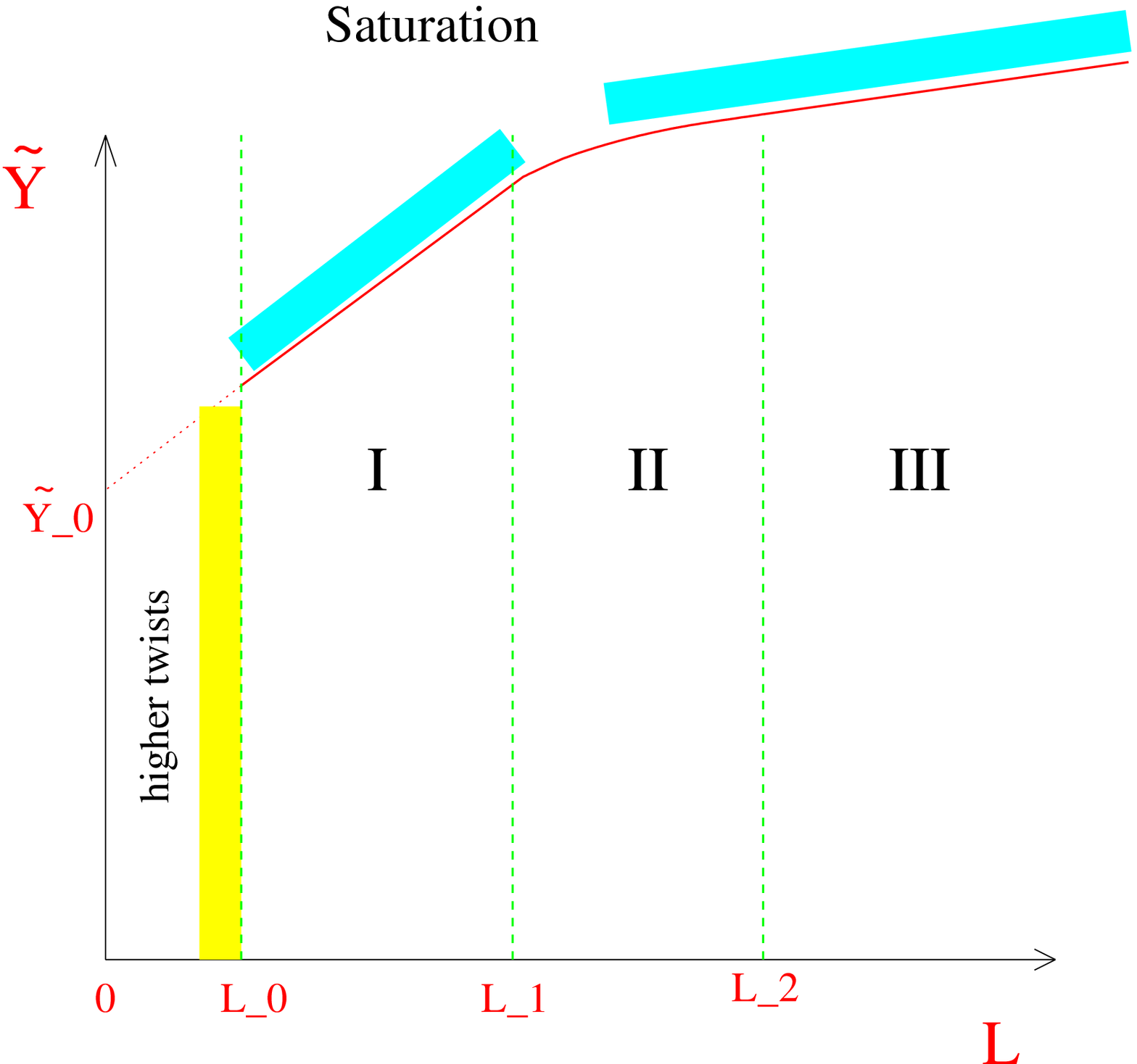}}} {\bf Figure 3}

{\it Critical Saturation line}

The critical saturation line is represented in the plane $\tY,L.$ The saturation 
region is under the curve. The region beyond the validity of leading twist 
approximation is below a value $1<<L_0.$ The regions {\bf I, II, III} correspond 
to different regimes of compensation between tree and one-loop dipole 
contributions (see text).

\end{document}